\documentclass[12pt]{article}
\usepackage{amssymb,epsf,graphicx}
\newcommand{\be}{\begin{equation}}
\newcommand{\ee}{\end{equation}}
\newcommand{\bea}{\begin{eqnarray}}
\newcommand{\eea}{\end{eqnarray}}
\newcommand{\bdm}{\begin{displaymath}}
\newcommand{\edm}{\end{displaymath}}

\def\<{\langle}
\def\>{\rangle}

\def\a{\alpha}
\def\b{\beta}

\def\d{\delta}
\def\g{\gamma}

\def\m{\mu}
\def\n{\nu}
  
  \def\th{\theta}                  %     \vartheta
\def\r{\rho}                                     %     \varrho
\def\s{\sigma}                                   %     \varsigma
\def\t{\tau}

\def\G{\Gamma}

\def\cl{{\cal L}}

\def\cn{{\cal N}}

\def\cv{{\cal V}}

\def\half{{1 \over 2}}

\def\l{\left}                    % These last three macros I use very often
\def\r{\right}
\let\eps = \varepsilon

\begin{document}
\baselineskip=15.5pt
\pagestyle{plain}
\setcounter{page}{1}
%--------+---------+---------+---------+---------+---------+---------+
%Body

%\begin{document}
\begin{flushright}
UW/PT 05-15\\
{\tt hep-th/0506265}
\end{flushright}

\vskip 2cm

\begin{center}
{\Large \bf Super Janus}
\vskip 1cm

{\bf A.~B.~Clark$^1$, A.~Karch$^1$} \\
\vskip 0.5cm
{\it  $^1$ Department of Physics, University of Washington, \\
Seattle, WA 98195-1560 \\}
{\tt  E-mail: abc@u.washington.edu, karch@phys.washington.edu} \\
\medskip

\end{center}

\vskip1cm

\begin{center}
{\bf Abstract}
\end{center}
\medskip

We propose and study a supersymmetric version of the Janus domain wall
solution of type IIB supergravity. Janus is dual to ${\cal N} =4$ super Yang
Mills theory with a coupling constant that jumps across an interface.
While the interface in the Janus field theory completely breaks all 
supersymmetries, it was found earlier that some supersymmetry can be restored
in the field theory at the cost of breaking the $SO(6)$ R-symmetry down
to at least $SU(3)$. We find the gravity dual to this supersymmetric
interface theory by studying the SU(3) invariant subsector of ${\cal N}=8$
gauged supergravity in 5d, which is described
by  
5D $\cn=2$ gauged supergravity with one hypermultiplet. 

\newpage

\section{Introduction}
\setcounter{equation}{0}
Recently, a domain wall solution to type IIB supergravity was
discovered which explicitly breaks supersymmetry while 
retaining stability \cite{janus}.  This solution, dubbed ``Janus''
after the Roman god with two faces,
includes a dilaton and a 5-form.  The dilaton 
develops a kink profile in this solution, approaching different
boundary values on either side of the domain wall (hence, two faces). 
Non-perturbative stability was proven for Janus-type solutions in
AdS$_d$ in \cite{fakesg}. 
This extends to IIB with the mild assumption that KK modes on
the $S_5$ factor of the geometry do not destabilize the solution.
In \cite{janusdual}, 
a gauge theory was proposed and investigated as an AdS/CFT dual of
this solution.  This gauge theory is ordinary $\cn=4$ 
Yang-Mills with different gauge couplings on either side of a defect
(or interface) that the boundary theory inherits from the 
bulk domain wall.  It was found that in the field theory, partial SUSY
could be restored by 
the addition of certain counter-terms localized to 
the defect at the cost of breaking the R-symmetry at least as far as SU(3).

We now propose a supergravity dual of this field theory, which we dub
Super Janus.  The uplift 
formulas of \cite{pilchwarner} allow one to obtain solutions of IIB SUGRA
in 10D from solutions of $\cn=8$ gauged SUGRA in 5D.  We restrict
ourselves to the subsector of this theory which is invariant under the
SU(3) R-symmetry of the field theory.  Under this restriction we find
4 scalars.  This subsector is well described by gauged $\cn=2$ SUGRA with one hypermultiplet,
where there is well-known formalism for solving the equations of
motion for supersymmetric domain walls \cite{bc,cdkv, cdl}.

\section{Review of Janus}
\subsection{Janus itself}
The Janus ansatz is an AdS$_4$-sliced domain wall supported by a
dilaton and 5-form: 
\bea
ds^2_{10} &=& e^{2U(\m)} \l( g_{ij} dx^i dx^j + d\m^2\r) + ds^2_{S^5} \\
\phi &=& \phi(\m) \\
F_5 &=& 4 \l( e^{5 U(\m)} d\m \wedge \omega_{AdS_4} + \omega_{S^5} \r),
\eea 
where $g_{ij}$ is the metric of AdS$_4$ with unit scale in an
arbitrary slicing.  With this ansatz, the IIB SUGRA equations of motion can
be simplified to the following two equations: 
\bea
\label{eq:janusdil}
\phi'(\mu) &=& c\ e^{-3 U(\m)} \\
\label{eq:januswarp}
U'(\mu) &=& \sqrt{e^{2 U} -1 + \frac{c^2}{24}\ e^{-6 U}},
\eea
where the constant, $c$, can be thought of as an (arbitrary)
integration constant arising from the dilaton's second 
order equation of motion:
\be
\partial_\m \l( \sqrt{g} g^{\m\n} \partial_\n \phi \r) =0.
\ee
The largest root (in terms of $U$) of the equation $U'(\m) =0$ then
determines the value of the warp factor on the domain wall, $U_0$, and
the range of the angular 
coordinate $\m$ is determined 
by integrating equation (\ref{eq:januswarp}):
\be
\label{eq:janusbdy}
\m_0 = \int_{U(0)}^\infty dU \sqrt{e^{2 U} -1 + \frac{c^2}{24}\ e^{-6 U}}.
\ee
Note that undeformed $AdS_5$ has $\m_0 = \pi/2$ while Janus has 
$\m_0 > \pi/2$.  There is a critical value of $c$ above which the
geometry becomes singular: 
\be
\label{eq:c-crit}
c_{cr} = \frac{9}{4 \sqrt{2}}\ .
\ee
For $c \geq c_{cr}$, the zeros (in $U$) of equation (\ref{eq:januswarp}) become
complex.  When this happens, 
it becomes impossible for the warp factor to have a turning point at
the wall.  Instead, the warp factor must be allowed to grow
arbitrarily negative as you approach the wall, and the geometry 
becomes nakedly singular.  This is discussed in
greater detail in \cite{janus, fakesg, janusdual, skenderis}.

An alternative interpretation is that one is free to pick any 
negative value for $U_0$ (within a range we will see shortly).  The constant 
$c$ is then determined by the vanishing of equation
(\ref{eq:januswarp}):
\be
c = \sqrt{24 e^{6U_0}(1 - e^{2U_0})},
\ee 
and the value of $\m_0$ is  
determined as before.  From this perspective, as we decrease $U_0$
from zero, we find that 
$c$ increases rapidly from zero to a maximum of $c_{cr}$ at
$U_0=-\ln \l(\frac{2}{\sqrt{3}}\r)$.  We must stop there, as we are
interested in the \emph{largest} root of equation (\ref{eq:januswarp}).
It would not do for the warp factor to have two turning points, so we
must restrict our choice to $U_0 \in \l(-\ln
  \l(\frac{2}{\sqrt{3}}\r),0\r)$.  While 
choosing the value of $c$ directly is  
more convenient for the study of  
Janus and its dual, when studying Super Janus it becomes more
convenient to specify the behavior of the warp 
factor on the domain wall (for numerical integration purposes).  We
will find Super Janus possesses an even narrower critical range for the values
of $U_0$.  The nature of this critical range of parameters in
Super Janus appears to be somewhat different from the critical range
we find in Janus.  It is also interesting to note that there is no
obvious interpretation of this criticality in the dual of either
theory.  
In a gauge theory with jumping coupling, there is no
obvious reason why the absolute amount of the jump in coupling
strength should cause a breakdown of any of the salient features
required for AdS/CFT duality.  We postpone deeper study of this for
the future.

\subsection{The Gauge theory dual of Janus}

The boundary theory dual to Janus was constructed and examined in
detail in \cite{janusdual}.  The essential point is that the
Lagrangian is simply that of $\cn=4$ Yang-Mills with a jump in the
gauge coupling at the defect inherited from the bulk.  Spatial
dependence of the gauge coupling means integration by parts is no
longer trivial, and consistency requires that one write the scalar
field kinetic term as $\half X^I D^\m D_\m X^I$, where $D_\m  X^I=
\partial_\m X^I + i[A_\m, X^I]$.  If we denote by $\cl'$ the
Lagrangian for $\cn=4$ Yang-Mills with this modified scalar kinetic
term normalized with the gauge coupling appearing only as $1/ g^2$, we
may write the action for the gauge theory dual of Janus as 
follows:
\be
S_{Janus} = \int d^4x \l( 1- \g \eps(x_3) \r) \cl'(x),
\ee
where $\eps$ is an odd step function, $\cl'$ is understood to have the
average gauge coupling ($g^{-2}=\bar{g}^{-2} = \half(g_+^{-2} +
g_-^{-2}$), and  
\be
\g = \frac{g^2_+ - g^2_-}{g^2_+ + g^2_-}.
\ee
The $\pm$ subscripts refer to the sign of $x_3$, i.e. on which side of
the defect one is located.  

\subsection{Restoring SUSY to the field theory dual of Janus}
In this subsection we briefly review the construction of partial
interface SUSY from appendix A of \cite{janusdual}.  This is the field
theory for which we will then construct a gravitational dual.  In
\cite{janusdual} this construction was used as a proof that it was not
possible for partial supersymmetry to creep back into the boundary
gauge theory dual of the explicitly non-supersymmetric Janus
solution.  The construction does, however, produce a perfectly good,
if somewhat peculiar, gauge theory.  

The first step is to consider the $\cn=4$ fields as being made up of
one $\cn=1$ vector multiplet and 3 $\cn=1$ chiral multiplets.  One
then considers how the SUSY variations of the $\cn=1$ Lagrangians are
modified by the inclusion of a Janus-style spatially varying coupling
constant (varying only in the $x_3$ direction which we will now denote
as $z$).  In practice, one treats the coupling as a continuous
function and takes the limit as it approaches a step function.  The
chiral Lagrangian we consider is 
\be
\cl = -\partial_\m \phi^*\partial^\mu \phi - \frac{i}{2} \bar{\psi}
\Gamma^\m \partial_\m \psi + F^*F + W'F - \frac{i}{2} W'' \bar{\psi}
P_+ \psi,
\ee
where $P_\pm=(1\pm \Gamma^5)$, and all dependence on the spatially
varying coupling constant is assumed to be in the superpotential,
$W$.  When considering the SUSY variation of this Lagrangian, one
finds that there are now terms which cannot be written as a total
derivative:
\be
\label{eq:chiralsusyvar}
\d \cl \supset -i\sqrt{2}\, \partial_z g \, \bar\eps \l(P_+ \Gamma^z
\frac{\delta {W'}^*}{\delta g} + P_- \Gamma^z \frac{\delta
W'}{\delta g} \r) \psi.
\ee
Now suppose that we preserve 2 supercharges in the general
spirit of supersymmetric defect conformal field theories \cite{dcft}.
Using the projector condition $\Pi \eps = \eps$, where
\be
\label{eq:martinsprojector}
\Pi = \frac{1+i\Gamma^5 \Gamma^z}{2},
\ee
the offending terms in $\d\cl$ (\ref{eq:chiralsusyvar}) can be rewritten as 
\be
\sqrt{2}\, \partial_z g\, \bar{\eps} \l(- P_-\frac{\d W'^*}{\d g}
  +P_+\frac{\d W'}{\d g} \r) \psi.
\ee
This expression happens to be a SUSY variation of $2\, \partial_z g\, \mathrm{Im}
  \frac{\d W}{\d g}$.  Thus, we may restore partial susy to the chiral
  sector by subtracting this term from the Lagrangian, effectively
  adding a counterterm.  This counterterm takes the form of a delta
  function in the limit where $g$ becomes a true step function, so we
  say the counterterm is localized on the defect.  Interestingly, this
  prescription fails if one normalizes the Lagrangian such that the
  coupling appears as an overall $1/g^2$ or if one writes the scalar
  kinetic term as $\phi^* \square \phi$.  

For the vector multiplet, the analogous prescription fails for the
``opposite'' normalization where we
normalize to put $g$ in the numerator.  We are forced to normalize the
vector Lagrangian as follows:
\be
\cl = -{1\over 4\, g^2} F_{\m\n}^a F^{a\m\n} - {i\over 2\, g^2} \bar{\lambda}^a
\G^\m D_\m \lambda^a + {1\over 2\, g^2} D^a D^a.
\ee
The offending terms in the SUSY variation of this Lagrangian may be
written as a supersymmetric variation of $\partial_z \l(
\frac{1}{4g^2} \r) \bar{\lambda}^a \G^5 \lambda^a$, after again
imposing the projector condition $\Pi \eps=\eps$, with $\Pi$ given in
(\ref{eq:martinsprojector}).  

Attempting to reassemble these Lagrangians into an $\cn=4$ multiplet
one finds no further obstruction to supersymmetry from gauge covariant
derivatives and Yukawa interactions.  However, the counterterms treat
the multiplets very differently, and, thus, the $\cn=1$ gaugino can no
longer be mixed with the fermions from the chiral multiplet.
Therefore, the maximal R-symmetry such a theory may possess is SU(3).
In the next section we construct the gravity dual of the theory with
this maximal R-symmetry.  Note that the boundary gauge theory only has
two supercharges but still possesses defect conformal symmetry, so the
SUSY preserved by the dual gravity theory will be $\cn=1$.  

\section{$\cn=8$ gauged SUGRA in 5D}
\label{section:sugra}
The $\cn=8$ gauged SUGRA theory has 42 scalars which comprise an
$E_{6(6)}/USp(8)$ coset.  These 42 scalars can be organized in
terms of the SO(6) R-symmetry of $\cn=4$ Yang-Mills by looking at the
symmetric traceless component of various operators.  Scalar masses
provide $6\times 6 \rightarrow (20' + 1)_s + 15_a$ which gives us 20
traceless scalars.  Fermion masses provide $4\times 4 \rightarrow
10_s + 6_a$, and $\bar{4} \times \bar{4} \rightarrow \overline{10}_s
+ \bar{6}_a$, which gives us a 10 and $\overline{10}$, for 20 more
scalars. The axion and dilaton round out our set of 42.  Next, we
truncate this to the SU(3) invariant subsector.  This sector of the
theory has been extensively studied in \cite{pilchwarner,distlerzamora}.  
The breaking pattern of our scalar reps as SO(6)
breaks to SU(3) is
\bea
20' & \rightarrow&  6+\bar{6}+8 \\
10  &\rightarrow & 1+6+\bar{3} \\
\overline{10}  &\rightarrow & 1+\bar{6}+3, 
\eea
giving us an additional 2 scalars which are singlets under our R-symmetry.
Thus, we find a total of 4 SU(3) invariant scalars.

These 4 scalar fields live on an
$\mathrm{SU}(2,1)/\mathrm{SU}(2)\times \mathrm{U}(1)$ coset, and
thus they can be naturally assembled into the scalar
sector of the universal hypermultiplet of $\cn=2$, 5D
SUGRA.  For this theory, 
there exists plentiful machinery for solving the BPS equations
\cite{bc, cdkv, cdl}.  Additionally, the work of \cite{bc} showed that with only hypermultiplets present, a simple consistency condition guarantees that the BPS equations will
solve the equations of motion.  We present and solve numerically
a supersymmetric domain wall ansatz which is smooth and displays the
correct qualitative features to be a gravity dual of the
supersymmetric version of the Janus boundary field theory, namely
one scalar field develops a kink profile, and the warp factor
turns around
at the domain wall.

\subsection{$\cn$=2 Flow Equations}
The scalar manifold of the $\cn=2$ theory with a single hypermultiplet
has 8 isometries to gauge.  Which isometries are gauged determine a
triplet of SU(2) Killing pre-potentials which 
in turn determine the scalar potential and the flow equations for
the scalars.  The most general form of the 8 Killing vectors and the 8
corresponding pre-potentials are given in \cite{cdkv}.  Our theory
really lives in IIB in 10D, so we are forced to choose our gauging
very carefully in order to match the fixed points of the scalar
potential.  From the 10D perspective, we expect a 2D plane of critical
points where 
the 10D dilaton and axion can take any value, but the other scalars
are fixed.  Thus we are forced to
pick a gauging for the $\cn=2$ theory that will give us a plane of
fixed points.  As in the $\mathrm{SU}(2)\times \mathrm{U}(1)$
invariant subsector studied in \cite{cdkv}, this requirement uniquely
fixes the gauging (up to a global symmetry transformation).  Additionally, we may identify the scalars
parameterizing this plane as the 5D dilaton and axion.  Unfortunately,
the uplift formulas generally entangle the 5D dilaton/axion with the
metric \cite{pilchwarner} so that it is not straightforward to read
off 10D behavior from 
5D or to unambiguously identify 10D fields with 5D counterparts, even
though \cite{pilchwarner} in principle tells us how to uplift.  Though
complicated, the existence of an uplift is guaranteed by \cite{cdkv},
which details an $\cn=2$ truncation of the FGPW flow in $\cn=8$ 5D
SUGRA \cite{fgpw}.  The field content and gauge structure of our model
can be embedded in that truncation.  

We parameterize the universal
hypermultiplet with the 4 scalars $V,\s,r,\a$.  This choice can be
obtained from the standard parametrization used in \cite{cdkv} by
redefining their $\th,\t$ fields as $\th=r \sin\a, \t=r \cos\a$. The
metric of the scalar manifold is 
\be 
\label{eq:scalarmetric} 
ds^2 = \frac{1}{2 V^2}\ dV^2 + \frac{1}{2V^2}\ d\s^2 
- \frac{2 r^2}{V^2}\ d\s
d\a + \frac{2}{V}\ dr^2 + \frac{2 r^2}{V} \l( 1+\frac{r^2}{V} \r)
d\a^2. 
\ee

In the language of \cite{cdkv}, the Killing vectors of the
quaternionic manifold are reorganized into generators of SU(2,1).
Constants $\a_i$ denote coefficients of SU(2) generators, and $\b$
denotes the coefficient of the compact U(1) (we do not consider
non-compact gaugings).  With only a hypermultiplet, the
gauged isometry must live in this $\rm{SU(2)}\times\rm{U(1)}$ subgroup.  Our
model differs from the ``Toy model with only a hypermultiplet'' of
\cite{cdkv} only
in the slicing of the domain wall, so the gauge and potential
structure must be the same.  In order to match the fixed point
structure of the potential as discussed 
above, we gauge the isometry corresponding to the choice $\b=-\a_3,
\a_1=\a_2=0$.  If one chooses $\b \neq |\vec{\a}|$,
then the fixed point structure consists of an
isolated critical point, UV in nature if $|\b| < |\vec{\a}|$ and IR if
$|\b|> |\vec{\a}|$.  The $\b=-\a_3$ gauge choice corresponds to the constant
shift of our angular scalar field $\a \rightarrow \a +c$.  The
critical point of the superpotential occurs at $r=0$ with $V$ and $\s$
free to take on any value.  Thus, we identify $V$ and $\s$ as 
linear combinations of the 5D dilaton and axion.  
Since $\s$ never appears in the scalar potential, either at the fixed
point or away from it, we can specify $\s$ as the axion.  Furthermore, turning
on $r$ and $\a$ corresponds to turning on 3-form flux dual to the
SU(3) singlet operators coming from the $10, \bar{10}$, i.e. gaugino
mass terms (modulus dual to $r$, phase dual to $\a$).  

The pre-potential is written in terms 
of an SU(2) phase, $Q^s$, and a superpotential, $W$, in the
following way: 
\be 
\label{eq:prepot} 
P^r = \sqrt{\frac{3}{2}} W Q^r.
\ee 
Our gauge choice gives us the superpotential 
\be 
\label{eq:W} 
W= \l( 1 + \frac{r^2}{V} \r), 
\ee 
and the SU(2)
phase 
\be 
\label{eq:Q} 
Q^s = \frac{1}{V+r^2} \l( - 2 r \sqrt{V}
\sin\a, -2 r \sqrt{V} \cos\a, V - r^2\r). 
\ee
With this gauging and parameterization, the full scalar potential has
a very simple form:%
\be
\label{eq:scalarpotential}
\cv = -6 + \frac{3 r^4}{V^2} - \frac{3 r^2}{V}.
\ee

Our AdS-sliced domain wall ansatz is:
\bea 
\label{eq:spacemetric} 
ds^2 &=& e^{2 U(z)} ds^2_{AdS4} + dz^2,  \\
V &=& V(z) \\
\s &=& \s(z) \\
r &=& r(z) \\
\a &=& \a(z).
\eea 

We can now use the machinery of \cite{bc, cdkv, cdl} to calculate
the flow equations for the warp factor and the hyper-scalars from
the vanishing of the fermionic supersymmetry variations.  The function, 
\be
\label{eq:gamma}
\g = \sqrt{1 -\frac{\lambda^2 e^{-2 U}}{g^2 W^2}} ,
\ee
is often used to simply express the resulting equations.  Throughout
this paper we use conventions such 
that lower-case Greek indices ($\m,\n$) refer to bulk space-time,
analogous lower-case latin indices 
(m,n) refer to space-time coordinates along the domain wall, lower-case 
Latin indices from later in the alphabet (r,s...) refer to SU(2)
structure, and upper-case Latin indices  
refer to hyperscalars or the
associated quaternionic geometry.  Quantities expressed as an SU(2)
triplet (e.g. $P^r$) can be 
re-expressed as a $2\times2$ matrix in the usual way:
\be
P_i^{\ j} \equiv i (\s_r)_i^{\ \ j} P^r.
\ee  
For a BPS solution we
require the fermionic supersymmetry variations to vanish.  In the
following expressions, ${\cal D}_\m$ is the total covariant derivative
(including both gravity and gauge structure), $f^{iA}_X$  
is the quaternionic vielbein,
and $\cn^{iA} = \frac{\sqrt{6}}{4} f^{iA}_X K^X(q)$, where $K^X(q)$ is
the gauged Killing vector.  The fermionic SUSY variations are then
\bea
\label{eq:gravitino}
\d \psi_{\m i} &=& {\cal D}_\m\ \eps_i  -
 \frac{i}{\sqrt{6}}\ g\ \g_\m\ 
 P_i^{\ j} \eps_j, \\
\label{eq:hyperino}
\d \zeta^A &=& \frac{i}{2}\ f^{iA}_X \g^\m (\partial_\m q^X) \eps_i -
g \cn^{iA} \eps_i. 
\eea
We use the residual supersymmetry projector of \cite{cdl}:
\be
\label{eq:fullprojector}
i \g_5 \eps_i = [ A(r) Q_i^{\ j} + B(r) M_i^{\ j}] \eps_j,
\ee
where $M^r$ is an SU(2) phase which depends on the scalar fields and
is orthogonal to $Q^r$ (i.e. 
$Q^r M^r=0$).  This is the most general  ansatz for residual
supersymmetry consistent with  
an AdS-sliced domain wall.  It was found in \cite{cdl} that the
quantities $A(r), B(r)$ are constrained 
up to signs by consistency and integrability constraints.  There is an additional consistency constraint on the choice of $M^r$ that is derived in \cite{bc,fakeness}:
\be
\l[ \th, \nabla_z\, \th \r] = - \sqrt{\frac{2}{3}} g \l[ \th, P \r],
\ee
and it should be noted that the conventions of \cite{bc} set $g=\sqrt{3/2}$.  
Writing this matrix equation in terms of SU(2) triplets we find
\be
\label{eq:consistency}
\th^r \nabla_z\, \th^s \eps^{rst} \s^t =  - \sqrt{\frac{2}{3}}\ g\ \th^r P^s 
\eps^{rst} \s^t.
\ee
Once this consistency condition is satisfied, the BPS equations will 
tell us the evolution of the scalar fields and the geometry.  

Integrability of the gravitino variation condition along the wall
($\d\psi_m=0$) and transverse to the wall ($\d\psi_5=0$) give two
different expressions involving $U'$.  These may be solved for $U'$
and the previously unknown function $A$:
\bea
U' &=& \pm \g(z)\ |g W|\\
A  &=& \mp \g(z),
\eea
where the sign choice in these two equations is correlated.
Note that consistency of the projector equation (\ref{eq:projector})
determines the magnitude but does not fix the sign of the function $B$:
\be
B = \pm \sqrt{1-\g^2},
\ee
where this upper/lower sign choice is independent of that in $A$.

As long as we satisfy equation (\ref{eq:consistency}), vanishing of the 
hyperini variations will now give us the flow
equations of the hyper scalars, according to the formulas of
\cite{cdkv, cdl} (equivalent expressions can be found in \cite{bc} using
somewhat different language, and a nice summary and dictionary between the
two languages can be found in \cite{fakeness}).  We first write down the 
general form of the flow equations with an arbitrary projector, $M^r$.  We 
will later make a specific choice and prove its consistency for the Super Janus gauging.  

Let $q^X$ denote
the $X^{\mathrm{th}}$ hyper scalar, $g$ the gauge coupling, $R^t_{XY}$
the SU(2) curvature (see \cite{cdkv} for the curvature formulas).  The
general flow equations are 
\be
q'^X = 3 g \l( A\ g^{XY} + 2 B\ \eps^{rst} M^r Q^s R^{tXY} \r)
\partial_Y W.
\ee
The signs of the $A$ and $B$ functions can be determined by demanding
that the warp factor has a turning point at the domain wall (matching to
Janus) and that the Killing spinor equation be continuous
across the domain wall.  The turning point condition gives 
\bea
\g(0) &=& 0\\
A &=& -\mathrm{sgn}(z)\g.
\eea
Continuity of the Killing spinor tells us that $B$ does not flip sign
at the wall; thus, 
\be
B= \sqrt{1-\g^2}.
\ee
This procedure is analogous to that followed in \cite{janus,fakesg}
for the original Janus solution.  In those papers, geodesic
completeness of the space-time demanded that the warp factor be
analytically continued as an even function of $z$ to the other side of
the wall.  (N.B: In those papers, the warp factor was denoted 
by $A(z)$ rather than $U(z)$).  We believe that this symmetry of the
original Janus solution was somehow accidental, as Super Janus is slightly
asymmetric, and there is no trace of this symmetry in the dual of
either theory.  

Before fixing the projector phase, $M^r$, the complete set of flow
equations for our parameterization is given by the following:
\bea
\label{eq:Uflow}
U'(z) &=& \pm g W \g = \pm \sqrt{g^2 \l(1 +\frac{r^2}{V} \r)^2 - 
\lambda^2 e^{-2 U}} \\
V'(z) &=& 6 g\, r \Biggl[ \pm\ r \g - \frac{1}{V+r^2} \biggl(\sqrt{V} 
\sqrt{1-\g^2} \bigl( -2 M^3 r\sqrt{V} + \\
&& (M^2 \cos\a + M^1 \sin\a) (r^2 -V) \bigr) \biggr) \Biggr] \\
\label{eq:sigmaflow}
\s'(z) &=& \frac{3 g\, r \sqrt{(1-\g^2)}}{\sqrt{V} (r^2 +V)} \Biggl(
M^2(-3 r^4 + 5r^2 V + 2 V^2)\sin\a \\
&& + 2 \cos\a \l( M^1(r^2-V)^2
+ M^3 r^3 \sqrt{V} \sin\a \r) \Biggr) \\
r'(z) &=& 3 g\, r \Biggl(\mp \g - \frac{r
  \sqrt{1-\g^2}}{(r^2+V)  
\sqrt{V}} \biggl(-2 M^3 r\sqrt{V} \\
&&+ \l(M^2 \cos\a + M^1 \sin\a\r) (r^2-V) \biggr)\Biggr) \\
\label{eq:alphaflow}
\a'(z) &=& \frac{3 g\, r \sqrt{(1-\g^2)}}{2\sqrt{ V} (r^2+V)} 
\l( \l(2 M^1 \cos\a -3 M^2 \sin\a\r) (r^2-V) + M^3 r \sqrt{V} \sin 2\a\r).
\eea
The sign choices come from the projector function $A$, thus, we must
choose the upper sign for $z>0$ and the lower sign for $z<0$.  We
again emphasize that this 
is consistent and smooth because Janus-like solutions require 
$U'\sim \g \rightarrow 0$ as $z\rightarrow 0$, and all terms that flip
sign at the  
domain wall are linear in $\g$ and thus continuous.  

We follow the strategy of \cite{cdl}, first we pick $\a(0)=0$ as part of our 
initial conditions, then an easy guess for our projector phase is 
\be
\label{eq:projector}
M^s = (0, Q^3, -Q^2) = \frac{1}{V+r^2} (0,-r^2 +V, 2 r \sqrt{V} ).
\ee
Now we must check equation (\ref{eq:consistency}) using 
$\th^r = -\g Q^r + \sqrt{1-\g^2} M^r$.  Since the Pauli matrices are linearly 
independent, we may drop them and regard (\ref{eq:consistency}) as three 
separate equations.  Then, because $\th^1=P^1=0$, only the $t=1$ 
component is non-trivial.  The action of the covariant derivative on $\th^r$
may be written as
\be
\nabla_z \th^r = {q^X}' \nabla_X \th^r + U' \partial_U \th^r.
\ee
With $\a(0)=0$ and the projector (\ref{eq:projector}), we find 
$\a'=\s'=0,$ and 
\bea
\label{eq:Vflow}
V'(z) &=& 6g  \l( \pm r^2 \g + r \sqrt{V}\sqrt{1-\g^2} \r)\\
\label{eq:rflow}
r'(z) &=& 3g  \l( \mp r\ \g + \frac{r^2}{\sqrt{V}} 
\sqrt{1-\g^2} \r).
\eea
This greatly simplifies the task of checking consistency:
\bea
\th^2 \nabla_z \th^3 - \th^3 \nabla_z \th^2 = - \l(1+ \frac{r^2}{V}\r) \sqrt{1
-\g^2}\\
\th^2 P^3 - \th^3 P^2 = {1\over g} \sqrt{3\over2} \l(1+ \frac{r^2}{V}\r) 
\sqrt{1-\g^2},\\
\eea
so we see that (\ref{eq:consistency}) is satisfied.  Since $M^r$ has three components, and there are 3 independent consistency equations, 
this completely specifies the projector phase.  

We now have a system of three coupled, non-linear, ordinary differential 
equations 
for the warp factor and two running scalars (\ref{eq:Uflow},
\ref{eq:Vflow}, \ref{eq:rflow}).   
We will solve these numerically in the next subsection.  

If one were to choose purely the upper sign in equations (\ref{eq:Uflow},
\ref{eq:Vflow}, \ref{eq:rflow}), then one obtains the solution of
\cite{cdl}, which is nakedly singular at the domain wall.  Super Janus
avoids this by the requirement that the warp factor have a turning
point at the domain wall, which we enforce through a careful choice of
initial conditions.  We believe the initial conditions chosen in
\cite{cdl} were such that they guaranteed a curvature 
singularity.  Indeed, we will find that only a narrow range of
parameter space allows a turning point.  

\subsection{Numerics}
We enforce the turning point condition with a suitable choice of
initial condition for $V$.  Setting equation (\ref{eq:Uflow}) to zero
at $z=0$, we obtain 
\be
V(0) = \frac{g\, r^2(0)}{\pm \lambda\, e^{-U_0} -g}
\ee
The scalar field $V$ must be strictly positive\footnote{Strictly
  speaking, $V$ must be strictly of one sign.  In many theories, it
  plays the role of the volume of a Calabi-Yau manifold, so it is
  conventional to choose it positive.  See \cite{cdkv} for more details.},
so we must choose the
plus sign, as there is no positive value of $\lambda\, e^{-U(0)}$ that
gives positive $V$ when the minus sign is chosen.  With the plus sign
choice, we must still require $\lambda\, e^{-U_0} >g$
for positivity of $V$.  An additional constraint on the initial
condition for the warp factor comes from the
requirement that the turning point be a minimum:  $U''>0.$  We may
easily calculate $U''$ using the BPS equations to replace $r'$,
$V'$, and $U'$ as needed.  The result is a surprisingly simple expression: 
\be
U''(z) = \lambda^2 e^{-2U(z)} - \frac{6g^2\, r^4}{V^2} - \frac{6g^2\, r^2}{V}.
\ee
Imposing the boundary condition for $V$ gives us
\be
U''(0) = \lambda\, e^{-U_0} \l( 6g -5 \lambda e^{-U_0}\r)
\ee
Positivity of this expression requires $\lambda\, e^{-U(0)}
<\frac{6}{5}\, g$.  Both constraints together limit us to a very 
narrow band: 
\be
1 < \lambda\, e^{-U(0)} < \frac{6}{5}\, g.
\ee

We may now numerically integrate equations (\ref{eq:Uflow},
\ref{eq:Vflow}, \ref{eq:rflow}).  As long as our initial conditions
are within the critical range, we find the scalar $V$ develops a
profile very much like the 5D dilaton in Janus.  By adjusting $r(0)$
(arbitrarily) and $\lambda\, e^{-U_0}$ (within criticality), we may
adjust the average value of $V$ as well as the split between the two
different asymptotic values of $V$.  

We now plot the numerical results for a typical choice of parameters:
$\lambda\, e^{-U_0} = 1.02, r_0=0.25$.  
\begin{figure}[!tbp]
\centering
\includegraphics{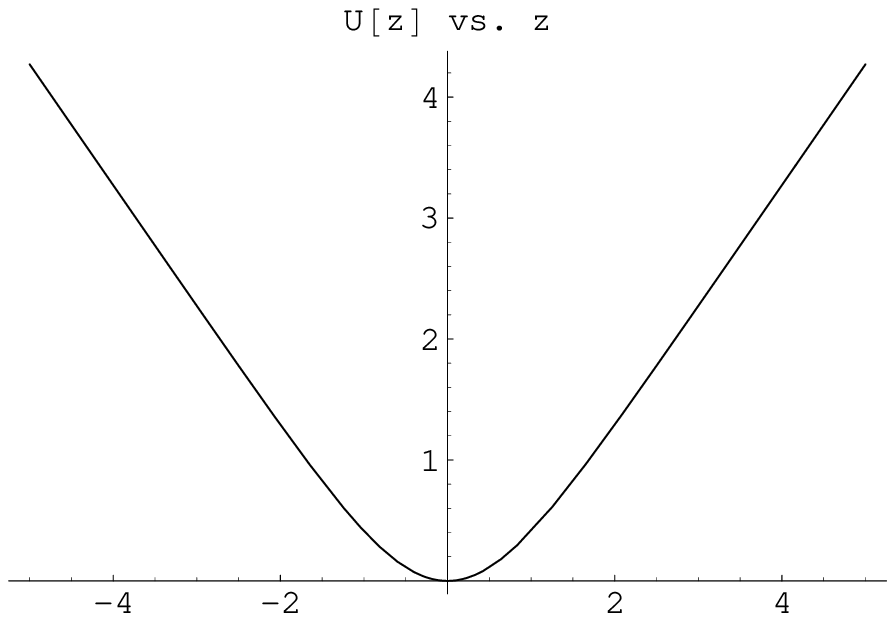}
\end{figure}
\begin{figure}[!tbp]
\centering
\includegraphics{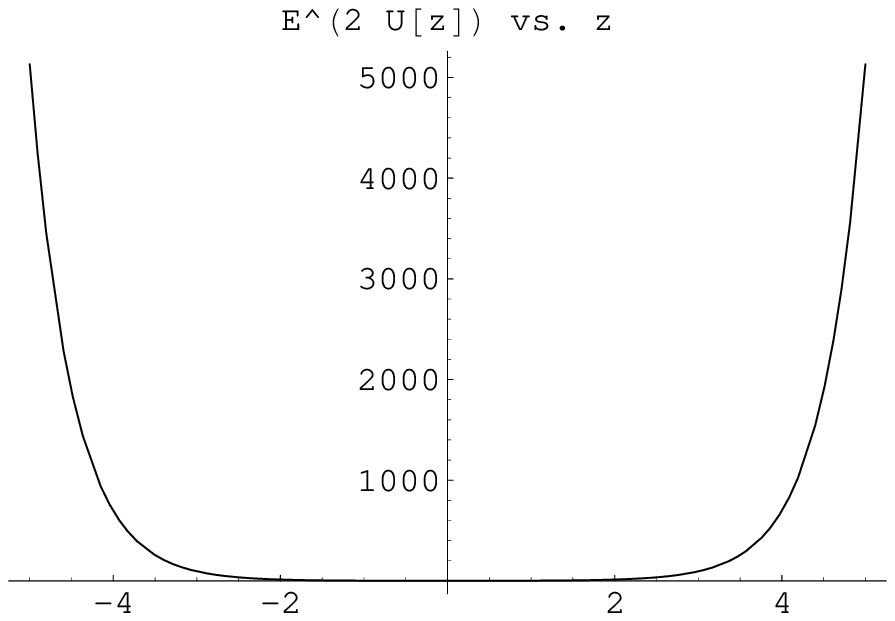}
\end{figure}
\begin{figure}[!tbp]
\centering
\includegraphics{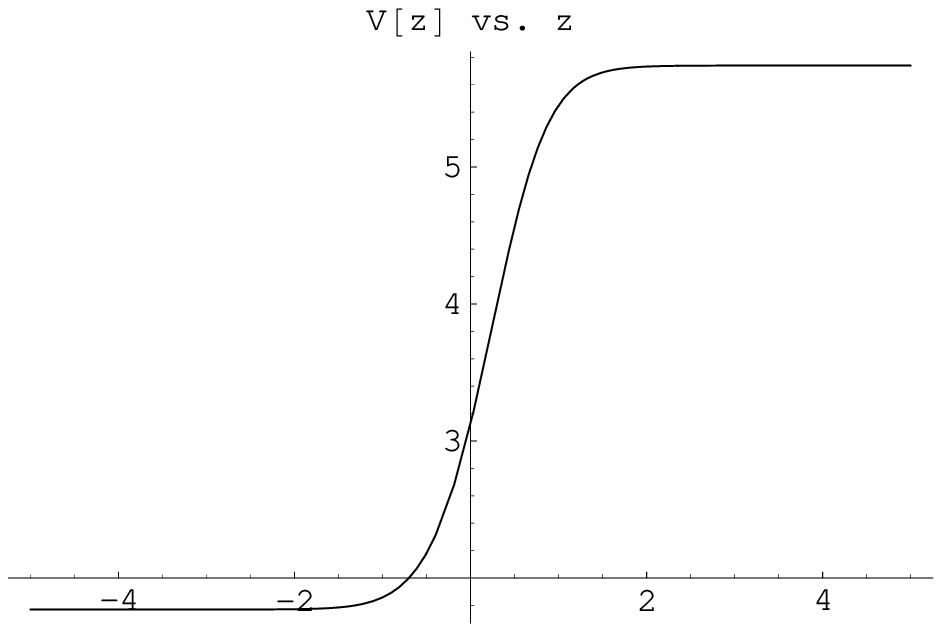}
\end{figure}
\begin{figure}[!tbp]
\centering
\includegraphics{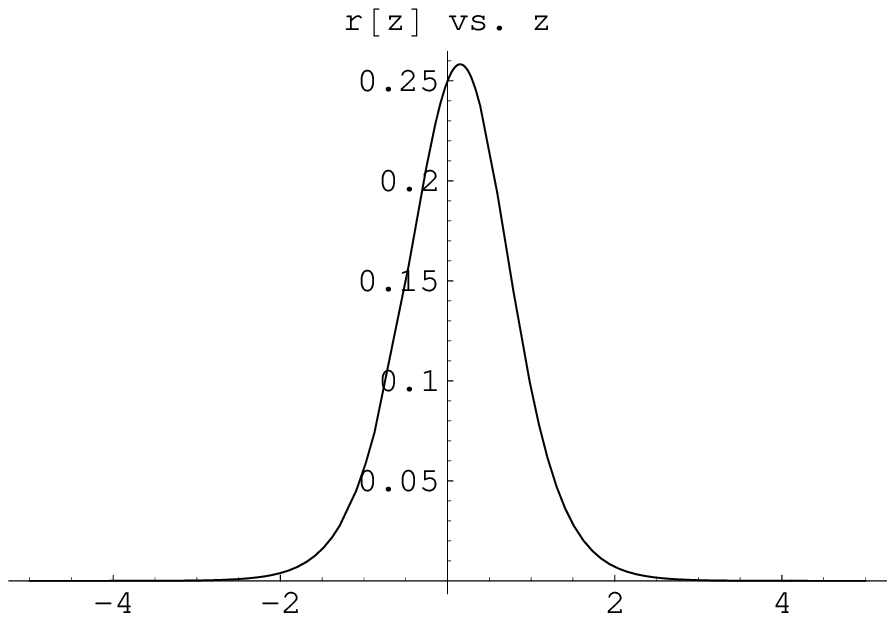}
\end{figure}

\subsection{Asymptotic behavior}
Using our knowledge from the numerics of the asymptotic behavior of
the fields $V$ and $r$ and the warp factor $U$, we can use the flow
equations (\ref{eq:Vflow},\ref{eq:rflow}) to determine the subleading
behavior for matching against the field theory dual.  The asymptotic
flow equations are 
\bea
\label{eq:Uasympt}
U'(z) & \mathop{\longrightarrow}\limits_{z \to \pm\infty} & \pm g\\
\label{eq:rasympt}
r'(z) & \mathop{\longrightarrow}\limits_{z \to \pm\infty} & \mp\ 3g\, r\\
\label{eq:Vasympt}
V'(z) &\mathop{\longrightarrow}\limits_{z \to \pm\infty}  & 6g \l( \pm r^2
+ r\sqrt{V} 
\frac{\lambda\, e^{-U}}{1 + \frac{r^2}{V}} \r).
\eea
Setting $g=1$ as we did in the numerics, equations (\ref{eq:Uasympt}, 
\ref{eq:rasympt}) may be easily solved to
yield
\bea
e^{U(z)} & \mathop{\longrightarrow}\limits_{z \to \pm\infty} & c_U \, 
e^{\pm z} \\
r(z) & \mathop{\longrightarrow}\limits_{z \to \pm\infty}  & c_r \, e^{\mp 3 z},
\eea
where we have left out a constant term in $r(z)$ because numerics show
$r \mathop{\longrightarrow}\limits_{z \to \pm\infty}  0$.  Plugging these
back into equation (\ref{eq:Vasympt}) and Taylor expanding in $e^{-z}$
we find
\be
V' \rightarrow 6 \l( c_r^2 \, e^{-6z}  + c_r \, c_U \lambda 
\sqrt{V} e^{-4z} \r),
\ee
which gives us
\be
V \rightarrow V_\pm - c_r \frac{3 \lambda\, c_U}{2} \sqrt{V_\pm} e^{-4z}.
\ee

This behavior is what we expect for $V$ being dual to a dimension 4
operator with source term and $r$ to a dimension 3 operator with no
source term, as expected if $V$ is the
5D dilaton and $r$ is related to 3-form flux (or gaugino mass terms).
In order to read off the vevs of the corresponding operators in the dual
gauge theory we need to determine the two integration constants
$c_U$ and $c_r$, which only get fixed by the IR behavior of the solution
and hence at the moment can only be determined from our numerical solution.

\section{Conclusion}
We have found a numerical solution for a supersymmetric domain wall in
5D, $\cn=2$ gauged supergravity which is supported by two
hyperscalars.  The BPS equations allow us to determine analytically
that the asymptotic behavior of these hyperscalars is appropriate for
the duals of a dimension 4 and a dimension 3 operator.  The uplift of
this solution to type IIB supergravity in 10D is the Super Janus
solution, dual to the gauge theory of \cite{janusdual} with partial
SUSY restoring counter-terms.  The 10D domain wall solution will be
supported by dilaton, 5-form flux, and 3-form flux dual to gaugino
mass.   

The existence of uplifts to 10D is guaranteed by the embedding in
\cite{cdkv} of the $\cn=2$ theory with a single hypermultiplet into
the FGPW solution \cite{fgpw} of the $\cn=8$ theory, where the uplift
formulas of \cite{pilchwarner} apply.  Unfortunately, to realize the uplift one
must understand how the hyperscalars of the $\cn=2$ theory sit inside
the 27-bein of the $E_{6(6)}/USp(8)$ coset of the $\cn=8$ theory,
which is far from 
straightforward.  Solving this would immediately give the uplift of
 Super Janus.  Another
potential route is to solve the 10D equations of motion directly with
some suitable ansatz deforming the AdS$_5\times S^5$ geometry to
something asymptotically AdS$_5$ crossed with an internal space
possessing only SU(3) isometry.  

It would also be interesting to understand the critical range of
parameters in Janus and Super Janus from the perspective of the
boundary gauge theories.  There is no obvious reason for the gauge
theory to care about the value of the jump in the coupling constant.
Finding the importance of this critical range in the gauge theory
would shed light on the Janus and Super Janus solutions.  

\section*{Acknowledgements}
We wish to thank D.~Freedman, M.~Schnabl, J.~Distler,
R.~Kallosh, M.~Zagerman, and E.~Silverstein for useful comments and
suggestions.  Special thanks to G.~Dall'Agata for comments on an earlier draft.  Additionally, ABC would like to thank R.~Van de Water
and A.~O'Bannon for advice on numerics and helpful discussions.  
Both ABC and AK are supported in part by DOE
contract \#DE-FG03-96-ER40956.


\begin{thebibliography}{99}

\bibitem{janus}
  D.~Bak, M.~Gutperle and S.~Hirano,
  %``A dilatonic deformation of AdS(5) and its field theory dual,''
  JHEP {\bf 0305}, 072 (2003)
  [arXiv:hep-th/0304129].

\bibitem{fakesg}
  D.~Z.~Freedman, C.~Nunez, M.~Schnabl and K.~Skenderis,
  %``Fake supergravity and domain wall stability,''
  Phys.\ Rev.\ D {\bf 69}, 104027 (2004)
  [arXiv:hep-th/0312055].
  
\bibitem{janusdual}
  A.~B.~Clark, D.~Z.~Freedman, A.~Karch and M.~Schnabl,
  %``The dual of Janus ((<:) <--> (:>)) an interface CFT,''
  Phys.\ Rev.\ D {\bf 71}, 066003 (2005)
  [arXiv:hep-th/0407073].

\bibitem{pilchwarner}
  K.~Pilch and N.~P.~Warner,
  %``N = 1 supersymmetric renormalization group flows from IIB supergravity,''
  Adv.\ Theor.\ Math.\ Phys.\  {\bf 4}, 627 (2002)
  [arXiv:hep-th/0006066].

\bibitem{skenderis}
  I.~Papadimitriou and K.~Skenderis,
  %``Correlation functions in holographic RG flows,''
  JHEP {\bf 0410}, 075 (2004)
  [arXiv:hep-th/0407071].

\bibitem{bc}
  K.~Behrndt and M.~Cvetic,
  %``Bent BPS domain walls of D = 5 N = 2 gauged supergravity coupled to
  %hypermultiplets,''
  Phys.\ Rev.\ D {\bf 65}, 126007 (2002)
  [arXiv:hep-th/0201272].

\bibitem{cdkv}
  A.~Ceresole, G.~Dall'Agata, R.~Kallosh and A.~Van Proeyen,
  %``Hypermultiplets, domain walls and supersymmetric attractors,''
  Phys.\ Rev.\ D {\bf 64}, 104006 (2001)
  [arXiv:hep-th/0104056].

\bibitem{cdl}
  G.~L.~Cardoso, G.~Dall'Agata and D.~Lust,
  %``Curved BPS domain walls and RG flow in five dimensions,''
  JHEP {\bf 0203}, 044 (2002)
  [arXiv:hep-th/0201270].

\bibitem{LopesCardoso:2001rt}
  G.~Lopes Cardoso, G.~Dall'Agata and D.~Lust,
  %``Curved BPS domain wall solutions in five-dimensional  gauged
  %supergravity,''
  JHEP {\bf 0107}, 026 (2001)
  [arXiv:hep-th/0104156].

\bibitem{Cardoso:2002ff}
  G.~L.~Cardoso and D.~Lust,
  %``The holographic RG flow in a field theory on a curved background,''
  JHEP {\bf 0209}, 028 (2002)
  [arXiv:hep-th/0207024].

\bibitem{Ceresole:2000jd}
  A.~Ceresole and G.~Dall'Agata,
  %``General matter coupled N = 2, D = 5 gauged supergravity,''
  Nucl.\ Phys.\ B {\bf 585}, 143 (2000)
  [arXiv:hep-th/0004111].

\bibitem{fakeness}
  A.~Celi, A.~Ceresole, G.~Dall'Agata, A.~Van Proeyen and M.~Zagermann,
  %``On the fakeness of fake supergravity,''
  Phys.\ Rev.\ D {\bf 71}, 045009 (2005)
  [arXiv:hep-th/0410126].

\bibitem{fgpw}
  D.~Z.~Freedman, S.~S.~Gubser, K.~Pilch and N.~P.~Warner,
  %``Renormalization group flows from holography supersymmetry and a
  %c-theorem,''
  Adv.\ Theor.\ Math.\ Phys.\  {\bf 3}, 363 (1999)
  [arXiv:hep-th/9904017].

\bibitem{dcft}
  O.~DeWolfe, D.~Z.~Freedman and H.~Ooguri,
  %``Holography and defect conformal field theories,''
  Phys.\ Rev.\ D {\bf 66}, 025009 (2002)
  [arXiv:hep-th/0111135].

% \bibitem{kklt}
%  S.~Kachru, R.~Kallosh, A.~Linde and S.~P.~Trivedi,
%  %``De Sitter vacua in string theory,''
%  Phys.\ Rev.\ D {\bf 68}, 046005 (2003)
%  [arXiv:hep-th/0301240].

\bibitem{distlerzamora}
  J.~Distler and F.~Zamora,
  %``Non-supersymmetric conformal field theories from stable anti-de Sitter
  %spaces,''
  Adv.\ Theor.\ Math.\ Phys.\  {\bf 2}, 1405 (1999)
  [arXiv:hep-th/9810206].

\end{thebibliography}
\end{document}